# Teaching Electrical Model of Power Transformers to Undergraduate Students: Magnetic Circuit Approach

Saeed Lotfifard, *Senior Member, IEEE*

*Abstract*— This paper explains a unified approach for teaching the electrical model of power transformers to undergraduate students using magnetic circuits. The commonly used approach for explaining the electrical model of power transformers is a hybrid approach in which magnetic circuits are used to explain the presence of series inductances. However, the presence of shunt inductance and resistance in the model is explained using alternative approaches. In contrary, this paper explains how both series and shunt elements can be described by using magnetic circuits. Moreover, three real-world examples and Matlab/Simulink results are provided to demonstrate how the presented explanations can be used to describe the responses of power transformers in real-world applications.

## I. INTRODUCTION

POWER transformers are an integral part of power systems. Therefore, understanding the model of power transformers is of special importance in power engineering undergraduate courses. The commonly used approach for explaining the electrical model of power transformers to the undergraduate students is based on a hybrid approach [1-3] in which magnetic circuits are used to explain the presence of series inductances in the electrical model. However, to explain the shunt elements different approaches are utilized. For instance, to develop the electrical model of real transformers in [4-5] the hysteresis loop is ignored. To explain the presence of both shunt resistance and inductance, the commonly used approach is based on decomposing the no-load current of the primary winding of the transformer into two waveforms. One of these waveforms is in phase with the applied voltage to the primary winding while the other one lags the applied voltage by ninety degrees. In this approach, it is discussed that to generate such waveforms, a parallel branch that includes a resistor and an inductor should be added to the electrical model.

While the above analysis is useful for explaining the behaviour of power transformers under the no-load condition, it is not consistent with the explanations that are presented for explaining the presence of the series inductance in the electrical model, which is based on the magnetic circuits analysis.

The objective of this paper is to explain a unified approach for describing the electrical model of power transformers

S. Lotfifard is with the school of Electrical Engineering and Computer Science, Washington State University, Pullman, WA 99164, USA (email: s.lotfifard@wsu.edu)

in which both series and shunt elements can be described using magnetic circuits. This approach provides direct connections between the real power transformer and corresponding elements in the electrical model using magnetic circuits.

The explained approach is based on the complex permeability [6-7]. It should be noted that different methods have been proposed for developing the electrical model of transformers core such as methods that are presented in [8-11]. However, the duality approach based on the complex permeability that is presented in [6-7] is easier to explain to the undergraduate students while it still has acceptable accuracy.

Three examples are provided that explain how the behavior of power transformers can be explained in real-world applications using magnetic circuits. Simulation results using Matlab/Simulink, which is a widely available software for studying electrical circuits [12], are also presented.

The rest of this paper is organized as follows: In section II a unified approach for explaining the electrical model of power transformers using magnetic circuits is presented. In Section III three examples of analyzing the behaviors of power transformers in real-world applications are presented. Finally, Section IV concludes the paper.

## II. MODELING TRANSFORMERS BASED ON MAGNETIC CIRCUITS

Consider a real power transformer (in contrast to ideal power transformers) shown in Fig.1. The copper losses which are resistive losses in the primary and secondary windings of the transformer are modeled by Rp and Rs in Fig. 2. The resistance of the high voltage winding is significantly larger than the resistance of the low voltage winding. This is because the low voltage winding has less number of turns and is made of wires with larger cross sectional area due to a higher nominal current value. For instance, for a typical transformer of 100 MVA, 7.97 kV:79.7 kV the resistance of the high voltage winding is 85 mΩ while the resistance of the low voltage winding is 0.76 mΩ [2].

According to Fig. 2, the flux in the transformer core at the primary winding side can be represented as follows:

$$\Phi_p = \Phi_{lp} + \Phi_m \qquad (1)$$

Where $\Phi_p$ is the primary flux, $\Phi_{lp}$ is the primary leakage flux, $\Phi_m$ is the flux linking both primary and secondary windings. In Fig.2, according to Faraday's law, the following holds:

$$V_p' = N_p \frac{d\Phi_p}{dt} = N_p \frac{d\Phi_{lp}}{dt} + N_p \frac{d\Phi_m}{dt} = V_{lp} + V_p'' \qquad (2)$$

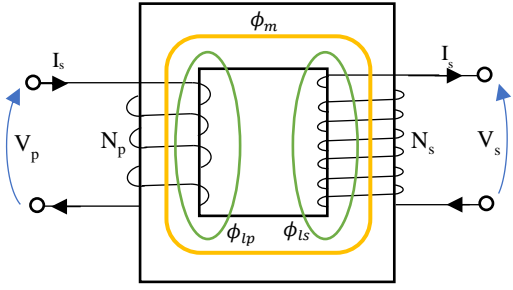

Fig.1 Schematic of a read power transformer

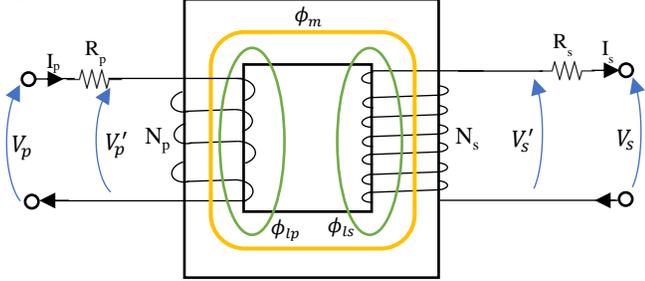

Fig.2 Schematic of a power transformer where the copper losses are modeled by Rp and Rs

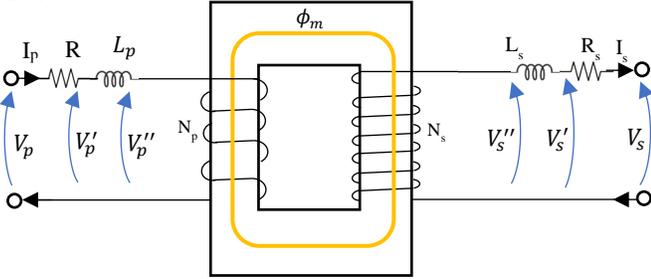

Fig.3 Schematic of a power transformer where the copper losses are modeled by Rp and Rs and the primary and secondary leakage fluxes are modeled by Lp and Ls

where $N_p$ is the number of turns in the primary winding, $N_p \frac{d\Phi_{lp}}{dt} = V_{lp}$ and $N_p \frac{d\Phi_m}{dt} = V_p''$. It is possible to write $\Phi_{lp} = \frac{N_p I_p}{\Re_{lp}}$ where $\Re_{lp}$ is the reluctance of the flux path through air.

Therefore, $V_{lp}$ can be represented as $\frac{N_p^2}{\Re_{lp}} \frac{dI_p}{dt} = V_{lp}$. Assuming $L_p = \frac{N_p^2}{\Re_{lp}}$, $V_{lp}$ can be represented as follows:

$$L_p \frac{dI_p}{dt} = V_{lp} \tag{3}$$

In the phasor domain (3) can be represented as follows:
$$L_p \omega j I_P = V_{LP} \tag{4}$$

The fact that in the phasor domain the derivative of a given sinusoidal signal can be represented by the multiplication of $\omega j$ and the phasor of that signal is used to derive (4) out of (3).

As the leakage inductance is proportional to the square of the number of winding turns, the leakage inductance of the high voltage side is significantly larger than the leakage inductance of the low voltage side of power transformers. For instance, for a typical 100 MVA, 7.97 kV:79.7 kV power transformer the leakage inductance of the high voltage winding is 10 mH while the leakage inductance of the low voltage winding is 0.11 mH [2].

It should be noted that in ideal power transformers the leakage flux $\Phi_{lp}$ is zero. As $\Phi_{lp} = \frac{N_p I_p}{\Re_{lp}}$, this means $\Re_{lp} = \infty$. Therefore, in ideal transformers $L_p = \frac{N_p^2}{\Re_{lp}} = \frac{N_p^2}{\infty} = 0$.

In Fig.3, according to Faraday's law, $V_p'' = N_p \frac{d\Phi_m}{dt}$ and $V_s'' = N_s \frac{d\Phi_m}{dt}$. Therefore, the following holds:

$$\frac{V_S''}{V_P''} = \frac{N_s}{N_p} \tag{5}$$

Further analysis of the electrical model of power transformers depends on the magnetic characteristic of the transformer core. The following two cases are considered:

*Case 1: Hysteresis loop of transformer core is ignored*
In this case it is assumed the hysteresis loop of B-H curve of the transformer core is ignored as shown in Fig.4-a.

The equivalent magnetic circuit is presented in Fig.5-a where $\Re = \frac{l}{\mu A}$ is the reluctance of the transformer core and A is the cross-sectional area of the transformer core, and $\mu$ is the permeability of the transformer core. By definition, $\mu = \frac{B}{H}$ which means $\mu$ at a given operating point of the transformer core is defined as the slope of a straight line from the origin to that given operating point on the B-H curve. For instance, Fig. 6 shows how the permeability can be calculated at two different operating points. As shown in Fig.6, the larger value of $\mu$ means for a given magnetic field intensity H (or a given applied current to the winding that is wrapped around the magnetic core), larger flux density B can be generated inside the magnetic core. By following the procedure shown in Fig. 6 for every operating point of B-H curve, $\mu$ can be calculated for a given B-H curve. For instance, Fig.7 [13] shows B-H curve of permalloy, and the corresponding relative permeability. It is a common practice to use relative permeability $\mu_r$ which is defined as $\mu_r = \frac{\mu}{\mu_0}$ where $\mu_0 = 4\pi 10^{-7}$H/m is the permeability of free space. As shown in Fig.7 $\mu$ is variable and its value depends on the operating point of the transformer core. $\mu$ has its maximum value at the knee point of the B-H curve. For the sake of simplicity, it is commonly assumed $\mu$ is a constant value during the non-saturated region of the magnetization curve. This means the non-saturated region of the B-H curve is approximated by a straight line that passes through the origin.

According to Fig.5-a, the following holds
$$N_p I_p - N_s I_s = \Re \Phi_m \tag{6}$$

In Fig. 3, according to Faraday's law, $V_p'' = N_p \frac{d\Phi_m}{dt}$. Using (6) the following holds:

$$V_p'' = N_p \frac{d\Phi_m}{dt} = \frac{N_p^2}{\Re} \frac{d}{dt}\left(I_p - \frac{N_s}{N_p} I_s\right) = \frac{N_p^2}{\Re} \frac{d}{dt}(I_{exc}) \tag{7}$$

Where $I_{exc} = \left(I_p - \frac{N_s}{N_p} I_s\right)$. Assuming $L_{exc} = \frac{N_p^2}{\Re}$, (7) can be represented as follows:

$$V_p'' = \frac{N_p^2}{\Re} \frac{d}{dt}(I_{exc}) = L_{exc} \frac{d}{dt}(I_{exc}) \tag{8}$$

In the phasor domain (8) can be represented as follows:
$$V_P'' = L_{exc} \omega j I_{EXC} = X_{exc} I_{EXC} \tag{9}$$

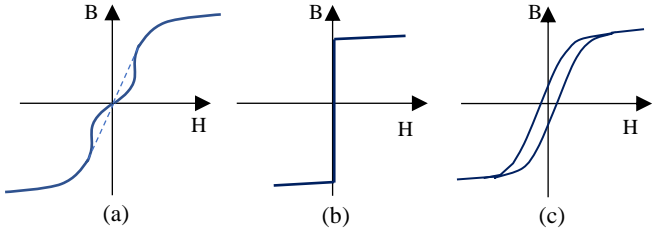

(a) (b) (c)

Fig.4 Magnetization curve of the transformer core (a) hysteresis loop is ignored, (b) ideal transformer (c) hysteresis loop is considered

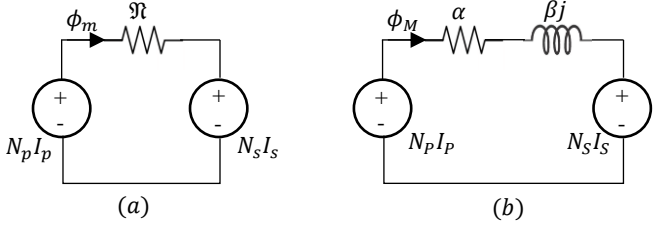

(a) (b)

Fig.5 (a) Magnetic circuit corresponding to (a) Case 1 where the hysteresis loop is ignored (b) Case 2 where the hysteresis loop is considered

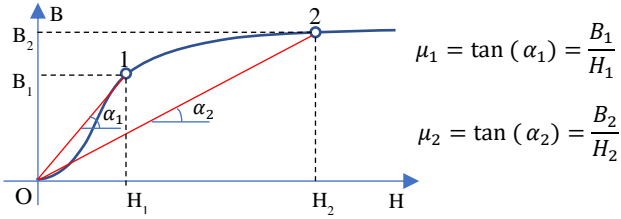

$$\mu_1 = \tan(\alpha_1) = \frac{B_1}{H_1}$$

$$\mu_2 = \tan(\alpha_2) = \frac{B_2}{H_2}$$

Fig.6 Schematic representation of the procedure for calculating the permeability at two different operating points

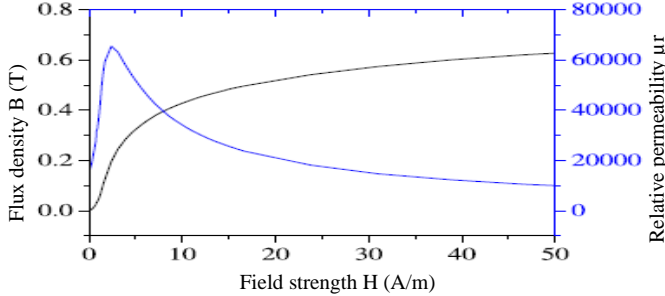

Fig. 7 Magnetic curve of permalloy and the corresponding relative permeability [13]

According to (5) and (8) and the fact that $I_{exc} = (I_p - \frac{N_s}{N_p} I_s)$, Fig. 3 can be modeled as Fig. 8. The impact of eddy current on the explained model in Fig.8 will be discussed later at the end of Case 2.

It should be noted that the shunt reactance $x_{exc}$ in Fig.8, is variable and its value depends on the operating point of the transformer. This is because $L_{exc} = \frac{N_p^2}{\Re} = \frac{N_p^2}{\frac{l}{A\mu}} = \frac{N_p^2}{l} A\mu$ and as $\mu$ is variable $L_{exc}$ is variable too.

The shunt element (i.e. the excitation branch) can be viewed as the representative of the required effort for circulating $\Phi_m$ in the transformer core. If the core of the transformer is considered to be ideal $\Re = 0$. Therefore, (6) becomes as follows:

$$N_p I_p - N_s I_s = 0 \qquad (10)$$

or

$$\frac{I_S}{I_P} = \frac{N_p}{N_s} \qquad (11)$$

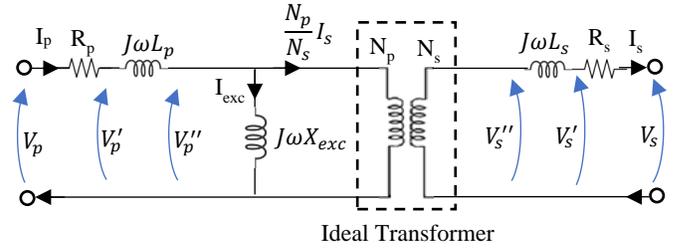

Fig.8 Schematic of the electrical model of power transformers corresponding to Case 1 where the hysteresis loop is ignored

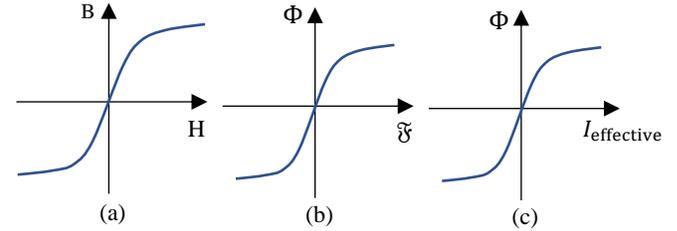

(a) (b) (c)

Fig.9 Magnetization curve of the transformer core in terms of (a) B and H, (b) Φ and 𝔉 (c) Φ and I

Therefore, according to (5) and (11), the shunt inductance in Fig. 8 should be removed. This means no effort is needed for circulating $\Phi_m$ inside the ideal transformer core.

Also as $\Re = \frac{l}{A\mu}$, $\Re = 0$ means $\mu = \infty$ in the non-saturated region of the transformer core. Therefore, B-H curve of an ideal transformer core looks like Fig. 4-b.

As shown in Fig.9-b, Φ-𝔉 curve has the similar shape as B-H curve. This is because for a given magnetic material the magnetic filed intensity (H) is directly related to the magneto-motive force (𝔉). The magnetic flux density (B) is also directly related to the magnetic flux (Φ). In Fig.9-b, magneto-motive force (𝔉) at a given section of a magnetic core is defined as the required efforts in terms of ampere-turns to drive the flow of the magnetic flux (Φ) in that given section of the magnetic core with the mean path length of $l_{section}$ and the cross sectional area of A. For example, if two windings are wrapped around the legs of a magnetic core similar to Fig. 3, the required effort to drive the flow of the magnetic flux of $\Phi_m = A \times B$ in the entire magnetic core with the mean path length of $l_c$ is $\mathfrak{F} = \oint H \cdot dl = H l_c$ which is equal to $N_p I_p - N_s I_s$.

It should be emphasised that at any given section of the magnetic core, the relationship between the flux (Φ) at that section and the magneto-motive force value (𝔉) at that section is defined by the Φ- 𝔉 curve shown in Fig.9-b. As in B-H curve, shown in Fig.9-a, at any point on the curve $\frac{B}{H} = \mu$, in Φ- 𝔉 curve, at any given point on the curve $\frac{\Phi}{\mathfrak{F}} = \frac{A \times B}{l_{section} \times H} = \frac{A}{l_{section}} \mu = \frac{1}{\frac{l_{section}}{A \times \mu}} = \frac{1}{\Re_{section}}$ where A is the cross-sectional area of the portion of the magnetic core for which Φ-𝔉 curve is defined and $l_{section}$ is the mean path length of the portion of the magnetic core for which the Φ- 𝔉 curve is defined. Therefore, Φ- 𝔉 curve of a given section of the magnetic core can be related to the B-H curve of the magnetic core by multiplying B-H curve by $\frac{A}{l_{section}}$.

In the case of a magnetic core similar to Fig.3, if the $\Phi$-$\mathfrak{F}$ curve of the entire magnetic core is provided, at any point on the $\Phi$-$\mathfrak{F}$ curve, $\frac{\Phi}{\mathfrak{F}} = \frac{A \times B}{l_c \times H} = \frac{A}{l_c} \mu = \frac{1}{\frac{l_c}{A \times \mu}} = \frac{1}{\mathfrak{R}_c}$ where $\mathfrak{R}_c$ represents the reluctance of the entire magnetic core and $l_c$ represents the mean path length of the entire magnetic core, and the horizontal axis of the $\Phi$-$\mathfrak{F}$ curve represents $\mathfrak{F}_{total} = N_p I_p - N_s I_s$.

Note that while B-H curve only depends on the type of the magnetic material, $\Phi$-$\mathfrak{F}$ curve in addition to the type of the magnetic material depends on the physical dimensions of the material for which $\Phi$-$\mathfrak{F}$ curve is defined. Therefore, for the same material with different dimensions the values of $\Phi$-$\mathfrak{F}$ curve are different.

As shown in Fig.9-c, it is also possible to use $\Phi$-I curve instead of $\Phi$-$\mathfrak{F}$ curve. To explain $I_{effective}$, in Fig.9-c, consider the magnetic core with two windings as shown in Fig.3. The total magneto-motive force can be defined as follows:

$$N_p I_p - N_s I_s = N_p \left(I_p - \frac{N_s}{N_p} I_s\right) = N_p I_{effective} \quad (12)$$

According to (12), if the current of $I_{effective} = I_p - \frac{N_s}{N_p} I_s$ is injected into the primary winding with $N_p$ turns, it creates the same magneto-motive force when currents of $I_p$ and $I_s$ are injected into two windings with $N_p$ and $N_s$ turns respectively. Note that, according to Fig.8, the effective current is equal to the excitation current as $I_{exc} = I_p - \frac{N_s}{N_p} I_s$. As an another example, assume three windings are wrapped around a magnetic core with a shape similar to Fig.3, and first and seconds windings generate magneto-motive forces with the same polarity while the third winding generate a magneto-motive force with the opposite polarity compared with other windings. In this case, $I_{effective}$ is determined as follows:

$$N_1 I_1 + N_2 I_2 - N_3 I_3 = N_1 \left(I_1 + \frac{N_2}{N_1} I_2 - \frac{N_3}{N_1} I_3\right) = N_1 I_{effective} \quad (13)$$

According to (13), $I_{effective} = I_1 + \frac{N_2}{N_1} I_2 - \frac{N_3}{N_1} I_3$

As in B-H curve, shown in Fig.9-a, at any point on the curve $\frac{B}{H} = \mu$, in $\Phi$-$I_{effective}$ curve of a magnetic core with the shape similar to Fig.3, at any point on the curve $\frac{\Phi}{I_{effective}} = \frac{A \times B}{\frac{l_c \times N_p \times I_{effective}}{N_p \times l_c}} = \frac{A \times B}{\frac{l_c \times H}{N_p}} = \frac{N_p \times A}{l_c} \times \frac{B}{H} = \frac{N_p \times A}{l_c} \mu$.

Therefore, $\Phi$-$I_{effective}$ curve of a magnetic core with the shape similar to Fig.3 can be related to the B-H curve by multiplying B-H curve by $\frac{N_p \times A}{l_c}$.

*Case 2: Hysteresis loop is considered*
In this case it is assumed the B-H curve of the transformer core is as shown in Fig.4-c. To model the presence of the hysteresis loop, the concept of complex permeability [6-7] can be utilized where the hysteresis loop is represented by an ellipse as shown in Fig.10. In Fig.11 the relationship between the B-H curve and B and H waveforms is shown schematically.

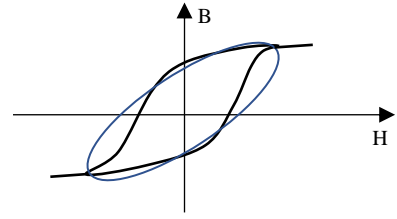

Fig.10 Representation of the hysteresis loop by an ellipse

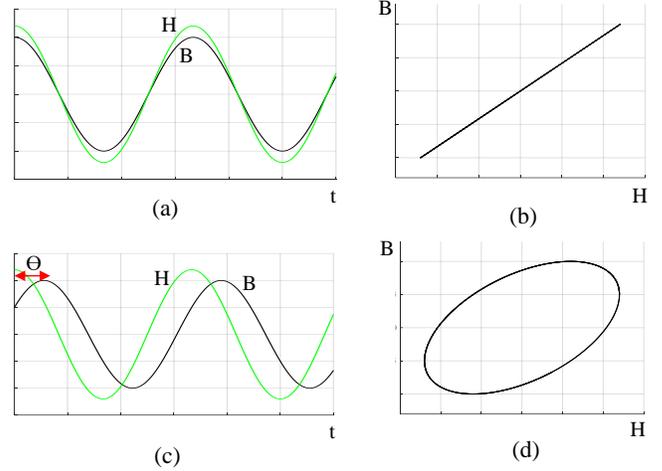

Fig.11 (a) B and H waveforms when there is no phase angle shifts between these two waveforms (b) B-H graph for the waveforms in (a), (c) B and H waveforms with a phase angle displacement (d) B-H graph for the waveforms in (c).

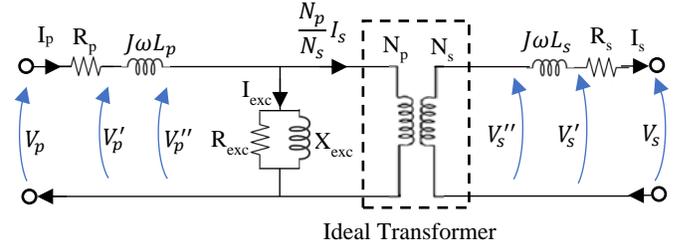

Fig.12 Schematic of the electrical model of power transformers corresponding to Case 2 where the hysteresis loop is considered

According to Fig.11-a and Fig.11-b, for sinusoidal B and H waveforms, if there is no phase angle shift between the waveforms, B-H graph is a straight line. According to Fig.11-c and Fig.11-d, if there is a phase angle shift between B and H waveforms B-H graph becomes an ellipse.

In Fig.11-c, B and H waveforms can be represented as phasors in the form of $B_{max} \angle -\theta$ and $H_{max} \angle 0$. Therefore, the following holds:

$$\mu = \frac{B}{H} = \frac{B_{max} \angle -\theta}{H_{max} \angle 0} = \frac{B_{max}}{H_{max}} \angle -\theta \quad (14)$$

According to (14), the difference between the permeability of the magnetization curve without the hysterics loop (which its permeability is a real value) and the permeability of the magnetization curve with hysterics loop is the presence of phase angle shift $-\theta$. This phase angle shift makes the permeability of the magnetic curve with hysteresis loop a complex value. Moreover, the following holds:

$$\frac{1}{\mu} = \frac{H}{B} = \frac{H_{max} \angle 0}{B_{max} \angle -\theta} = \frac{H_{max}}{B_{max}} \angle \theta = \alpha' + \beta' j \quad (15)$$

Where $\alpha' = \frac{H_{max}}{B_{max}} \cos(\theta)$ and $\beta' = \frac{H_{max}}{B_{max}} \sin(\theta)$. According to (15), the reluctance of the transformer core becomes as follows:

$$\Re = \frac{l}{A\mu} = \frac{l}{A} \times \frac{1}{\mu} = \frac{l}{A} \times (\alpha' + \beta'j) = \alpha + \beta j \quad (16)$$

Therefore, (6) in the phasor domain becomes as follows:
$$N_p I_P - N_s I_S = \Re \Phi_M = (\alpha + \beta j) \Phi_M \quad (17)$$

Therefore, the following holds:
$$\Phi_M = \frac{N_p I_P - N_s I_S}{\alpha + \beta j} \quad (18)$$

Moreover, in the phasor domain $V_p'' = N_p \frac{d\Phi_m}{dt}$ can be represented as follows:
$$V_P'' = N_p \omega j \Phi_M \quad (19)$$

Substituting (18) into (19), the following can be derived
$$V_P'' = N_p \omega j \frac{(N_P I_P - N_S I_S)}{(\alpha + \beta j)} = N_p^2 \omega j \frac{(I_P - \frac{N_S}{N_p} I_S)}{(\alpha + \beta j)} = N_p^2 \omega j \frac{I_{EXC}}{(\alpha + \beta j)} \quad (20)$$

Where $I_{EXC} = (I_P - \frac{N_s}{N_p} I_S)$ is the phasor of the current of the excitation branch.

According to (20), the following holds:
$$\frac{I_{EXC}}{V_P''} = \frac{(\alpha + \beta j)}{N_p^2 \omega j} = \frac{(\beta - \alpha j)}{N_p^2 \omega} = \frac{1}{\underbrace{\frac{N_p^2 \omega}{\beta}}_{}} + \frac{1}{\underbrace{\frac{N_p^2 \omega j}{\alpha}}_{}} = \frac{1}{R_{exc}} + \frac{1}{X_{exc}} \quad (21)$$

Therefore, according to (5) and (21) and the fact that $I_{EXC} = (I_P - \frac{N_s}{N_p} I_S)$, Fig.3 can be represented as Fig.12.

Comparing Fig.12 and Fig.8, it can be observed that $R_{exc}$ is added to the model when the hysteresis loop is considered. According to Fig. 12, the power loss in $R_{exc}$ can be calculated as $\frac{V_P''^2}{R_{exc}}$. This power loss represents the power loss related to the hysteresis loop due to the reorientation of the magnetic domains of the transformer core. More details on the physics of this power loss are presented in standard textbooks on electrical machines such as [1].

The model presented in Fig.12 is only valid for the non-saturated region of the magnetization curve. For the saturated region of the magnetization curve, the model in Fig. 8 is valid with $\mu$ of the saturated region of B-H curve.

Eddy current loss is another type of loss that was not discussed in developing models in Fig.8 and Fig. 12. Due to the time-varying flux inside the transformer core, voltage is induced in the transformer core. This induced voltage forms a flow of current in a circular pattern within the transformer core that looks like the eddies at a river bank. This circular current within the transformer core, causes a power loss. Similar to the hysteresis loss, eddy current loss, which is also corresponds to the loss within the transformer core, can be modeled by a resistance in parallel with the transformer core magnetization inductance in Fig.8. In Fig.12, the resistance that represents the hysteresis loss and the resistance that represents eddy current loss can be combined and modeled by a single resistance that represents the transformer core loss.

### III. EXAMPLES OF ANALYZING THE BEHAVIOR OF POWER TRANSFORMERS

The following three examples demonstrate how the presented explanations in the previous section can be used for describing the behaviour of transformers in real-world applications.

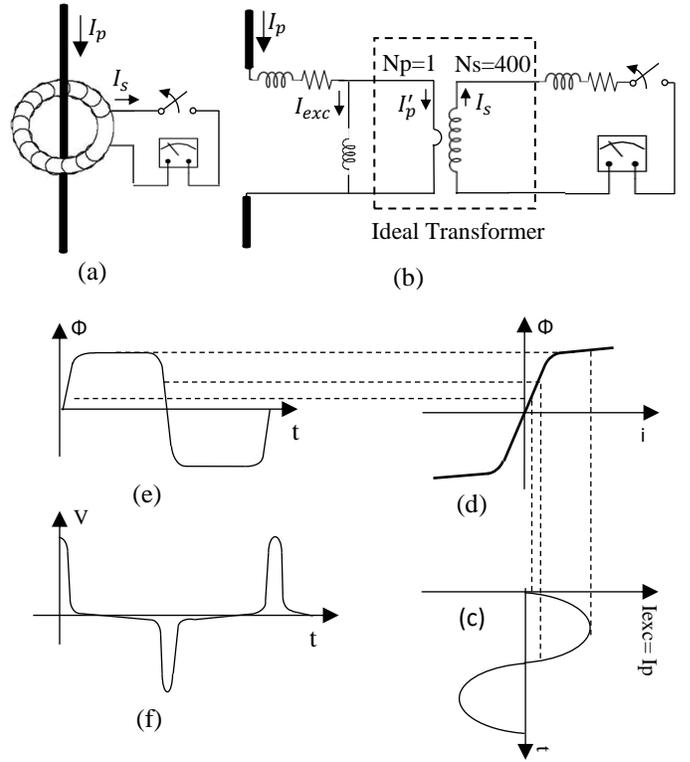

Fig.13 (a) schematic of a current transformer with opened secondary (b) equivalent electrical circuit of the current transformer with opened secondary (c) excitation current (d) magnetization curve of the CT (e) flux of the CT (f) voltage of the secondary cide of the CT when the secondary side is open

*Example 1: Current Transformers with Opened Secondary Side*
Current transformers (CT) reduce the current levels of the primary side to the acceptable levels at the secondary side to be injected into metering and/or protective devices. Fig.13-a shows a CT with 2000/5 ratio that is connected to an Amp meter. This CT is a transformer with one turn winding at the primary side and a winding of 400 turns at the secondary side. The secondary side of the CT must be short circuited before disconnecting the Amp meter. The followings explain what happens if the secondary side of the CT becomes open circuited. This example provides insights into the impact of transformer core B-H curve nonlinearity on the response of the transformer.

Assume the secondary side of the CT is open circuited. The voltage of the secondary side of CT can be determined according to the equivalent circuit of the transformer and the H-B curve (for the sake of simplicity, hysteresis loop is not considered).

Also shown in Fig.13, the current of the primary side of the CT is Ip. As shown in Fig.13-b, as the secondary side of the CT is open circuited, $I_p' = \frac{N_p}{N_s} I_s = 0$ which means the current of the excitation branch $I_{exc}$ is equal to Ip. Accordingly, Fig.13-e shows the flux of the current transformer core. The voltages of the secondary side of the CT can be calculated using $V_s'' = N_s \frac{d\Phi}{dt}$ as shown in Fig.13-f.

According to Fig.13-f, when the CT is saturated, voltage becomes almost zero which means $L_m$ reduces significantly (i.e. it is short circuited).

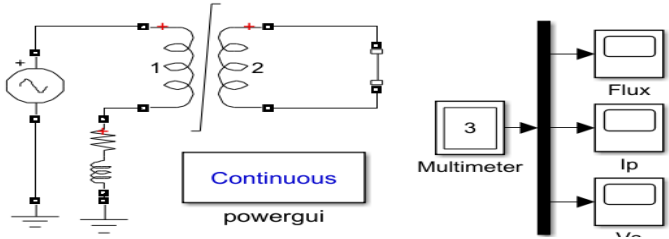

Fig.14 schematic of simulated circuit in Matlab/Simulink to study the behaviour of a current transformer when its secondary side is opened

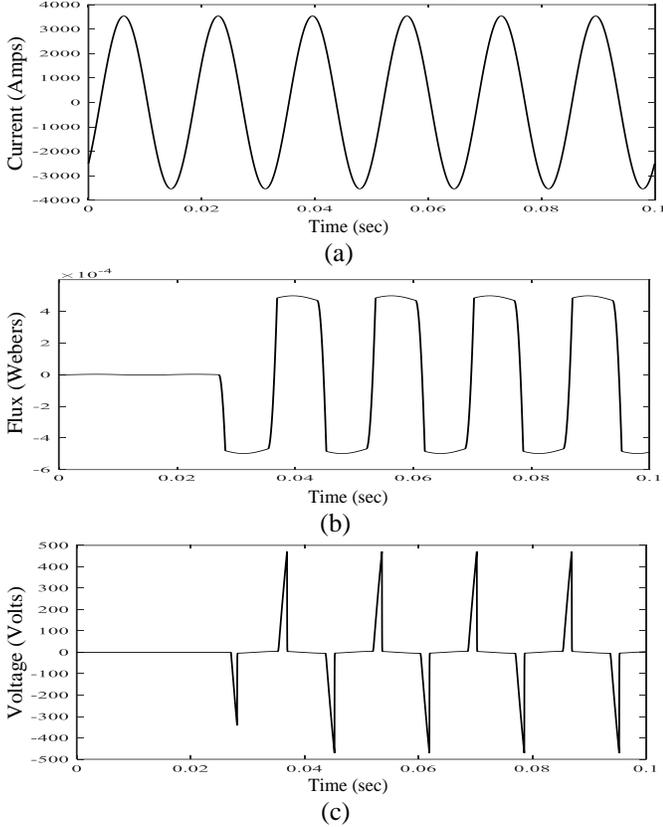

Fig.15 (a) Primary current (b) flux and (c) voltage of the secondary side of the CT when the secondary side is open circuited

This is because $L_{exc} = \frac{N_p^2}{\Re}$ where $\Re = \frac{l}{A\mu}$. During the saturation of the CT, $\mu$ reduces significantly, which means $\Re$ increases significantly and consequently $L_m$ reduces drastically.

Fig. 14 shows a simulated circuit in Matlab/Simulink that includes a single-phase AC source and a load of 1+1j ohms. The CT ratio is 2000/5. The secondary side of the CT is open circuited at t= 0.027 seconds. Fig.15 shows the primary current, flux, and the secondary voltage. The simulation results are consistent with the theoretical analyses presented in Fig.13.

*Example 2: Transformer Winding Resistance Measurement*

One of the common tests of power transformers is winding resistance measurement for the health monitoring of the transformer. A DC source is connected to the winding of the transformer and voltage and current of the winding is measured. Once the current is stabilized (i.e. reach the steady state condition), the resistance of the winding is calculated by

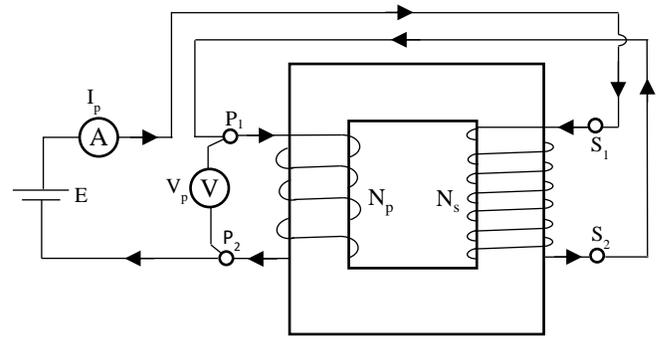

Fig.16 schematic of double approach for accelerated measurement of the transformer winding resistance dividing the measured voltage over the measured current. To reduce the test duration, the time constant of the circuit should be reduced. As will be discussed in the subsequent paragraphs, an effective approach for reducing the time constant is operating the transformer at the saturated region of the magnetization curve during the test.

The transformer core typically saturates for $I_{exc} = 1 - 5\%$ of the nominal current (i.e. Ip=1-5% of the nominal current when the secondary side is open circuited). Therefore, the minimum DC current of value of 1-5% of the transformer nominal current is selected for the test current to achieve transformer saturation. However, the test current should not exceed 10% of the transformer nominal current as it causes erroneous readings due to heating of the winding [14-15].

For large three-phase transformers, the winding resistance measurement of the low voltage winding of the transformer may take a long time. It may take up to 30 minutes for the current to stabilize [14-15] for each phase of the low voltage windings which prolongs the test duration. This is because due to the small number of turns in the low voltage winding, to generate enough magneto-motive force ($\mathfrak{F} = NI$) to saturates the transformer core, a higher current injection is needed. However, injecting such a high current may not be possible due to the limitation of the testing equipment and the need for larger cable. Moreover, to prevent the heating of the transformer winding that may cause erroneous readings, the injected current should not be more than 10% of the nominal current of the winding. A practical approach for saturating transformers with the acceptable current injections into the low voltage winding is using the double approach [14-15]. In this approach, as shown in Fig.16, the high voltage winding is connected in series with the primary side. The following explains describe the response of the transformer during the test.

In Fig. 8, assume a DC voltage of E is applied at the primary winding terminals while the secondary winding is open circuited. According to Fig.3 and Fig.16 and by ignoring $\phi_{lp}$ and $\phi_{ls}$ compared to $\Phi_m$ (i.e. assuming $L_p = L_s = 0$) the following holds:

$$E = (R_p + R_s)I_p + V_s^" + V_p^" = \left(\frac{R_s}{R_p} + 1\right)R_p I_p + \left(1 + \frac{N_s}{N_p}\right)V_p^" = \left(\frac{R_s}{R_p} + 1\right)R_p I_p + \left(\frac{N_s}{N_p} + 1\right)N_p \frac{d\Phi_m}{dt} \quad (22)$$

However, according to Fig.9-c, and assuming the slope of $\Phi - I_{effective}$ curve for the non-saturated region is m and for the saturated region is $m'$, the following holds

$$\begin{cases} \phi_m = mI_{effective} & \text{Non-saturated} \\ \phi_m = m'(I_{effective} - I^0_{effective}) - \phi^0_m & \text{Saturated} \end{cases} \quad (23)$$

Where $\phi^0_m$ and $I^0_{effective}$ are the flux and current values at the knee point of the $\Phi - I_{effective}$ curve. Therefore, the following holds:

$$\begin{cases} \dfrac{d\phi_m}{dt} = m\dfrac{dI_{effective}}{dt} & \text{Non-saturated} \\ \dfrac{d\phi_m}{dt} = m'\dfrac{dI_{effective}}{dt} & \text{Saturated} \end{cases} \quad (24)$$

According to (24), (22) can be represented as follows for Non-saturated region

$$E = \left(\dfrac{R_s}{R_p} + 1\right) R_p I_p + \left(\dfrac{N_s}{N_p} + 1\right) N_p m \dfrac{dI_{effective}}{dt} \quad (25)$$

As discussed before, in two winding transformers $I_{effective} = I_p - \dfrac{N_s}{N_p} I_s$ and as in the configuration of Fig.16, $I_p = -I_s$, for Fig.16 $I_{effective} = (1 + \dfrac{N_s}{N_p})I_p$. Therefore, (25) can be presented as follows:

$$E = \left(\dfrac{R_s}{R_p} + 1\right) R_p I_p + \left(\dfrac{N_s}{N_p} + 1\right)^2 N_p m \dfrac{dI_p}{dt} \quad (26)$$

Therefore, the time constant of the circuit in Fig.16, becomes $\tau = \dfrac{\left(\dfrac{N_s}{N_p}+1\right)^2}{\left(\dfrac{R_s}{R_p}+1\right)} \dfrac{N_p m}{R_p}$. Note that the increase in the value of the time constant due to $\left(\dfrac{N_s}{N_p} + 1\right)^2$ is canceled out by $(\dfrac{R_s}{R_p} + 1)$ as $R_s$ (i.e. the resistance of the high voltage side) is significantly larger than $R_p$ (i.e. the resistance of the low voltage side). Therefore, the time constant of the non-saturated region of Fig.16 is almost similar to the case if the source is only applied to the low voltage winding and the high voltage winding is open circuited. However, the advantage of the configuration of Fig.16 is that $I_{effective} = (1 + \dfrac{N_s}{N_p})I_p$ while for the case that the source is only applied to the low voltage winding and the high voltage winding is open circuited $I_{effective} = I_p$. Therefore, due to the configuration of Fig.16, higher $I_{effective}$ can be generated. This means the transformer core can be saturated by less Ip current. Once the transformer core is saturated, the time constant becomes $\tau' = \dfrac{\left(\dfrac{N_s}{N_p}+1\right)^2}{\left(\dfrac{R_s}{R_p}+1\right)} \dfrac{N_p m'}{R_p}$. As $m'$ is significantly smaller than m, $\tau'$ becomes significantly smaller than $\tau$ which means the current reaches the steady state condition faster compared with the case where the transformer is not saturated.

A power transformer is simulated in Matlab/Simulink and its terminals are connected to the source according to Fig. 16. In the simulations, for the sake of simplicity of the implementation and also to better demonstrate the impact of the configuration of Fig.16 on the time constant of the test system, it is assumed the voltage source is constant and its value is equal to the value of the resistance of the primary winding multiplied by the determined steady state current value.

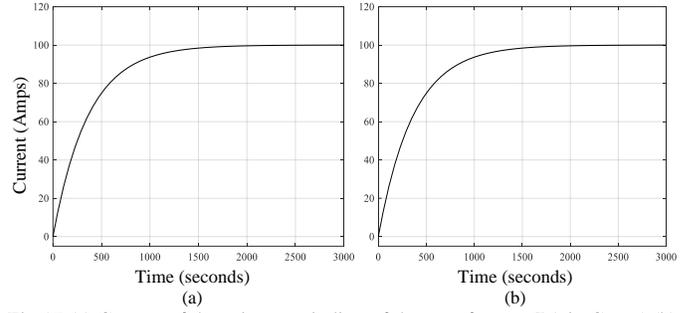

Fig.17 (a) Current of the primary winding of the transformer (Ip) in Case 1 (b) excitation current of the transformer (Iexc,) in Case 1

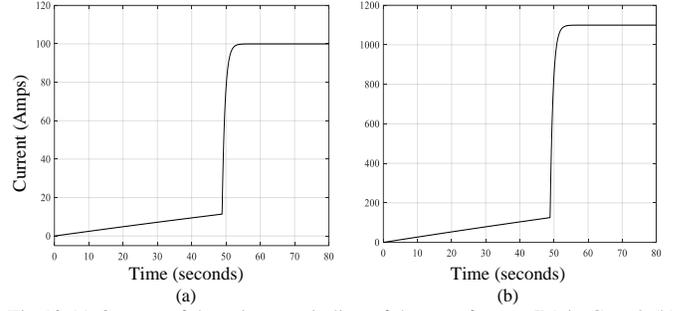

Fig.18 (a) Current of the primary winding of the transformer (Ip) in Case 2 (b) excitation current of the transformer (Iexc) in Case 2

Two cases are studied. In Case 1, a single phase power transformer of 100 MVA, 7.97 kV:79.7 kV with the resistance of 85 mΩ at high voltage winding and the resistance of 0.76 mΩ at the low voltage winding is selected. A voltage source is applied to the low voltage winding of the transformer such that at the steady state condition 100 amperes is injected into the transformer while the high voltage winding is open circuited. In Case 2, the same transformer is selected but the transformer terminals are connected to the source as shown in Fig.16. In this case the voltage source is set such that at the steady state condition the current of 100 amperes is injected into the transformer. Fig. 17 and Fig. 18 show the primary current of the power transformer and the excitation currents for both cases. As shown in these figures, the current stabilizes significantly faster in Case 2. Note that in Case 2, $I_{exc} = (1 + \dfrac{79.7}{7.97})I_p = 11 \times I_p$ while is Case 1 $I_{exc} = I_p$.

*Example 3: Peaking Transformers*
Peaking transformers are special transformers that can generate pulses from sinusoidal signals [16-17]. For instance, they used to be utilized for generating sharp peaked waveforms for firing thyratron tubes [16-17]. Analysing the operation of peaking transformers is a good exercise for a better understanding of characteristics of the magnetization curve of transformers core.

Fig. 19 shows the schematic of a peaking transformer. In this example Np=1000 turns, Ns=2000 turns, μr=100000. The core of the transformer saturates at 0.7 Tesla. Once a section of the transformer core is saturated, μr reduces significantly and the reluctance of the saturated section increases drastically. Therefore, it is assumed no further magnetic flux can flow through the saturated section.

The following sinusoidal voltage signal is applied at the primary winding of the transformer

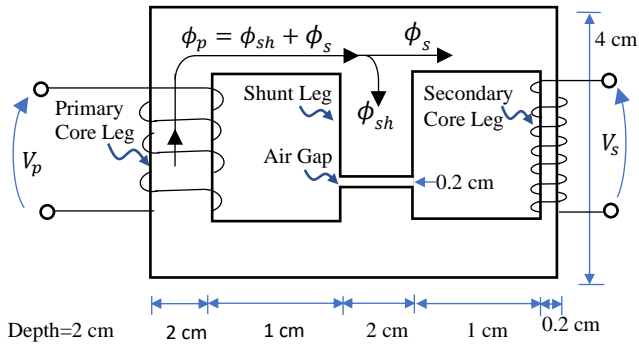

Fig. 19 Schematic of a peaking transformer

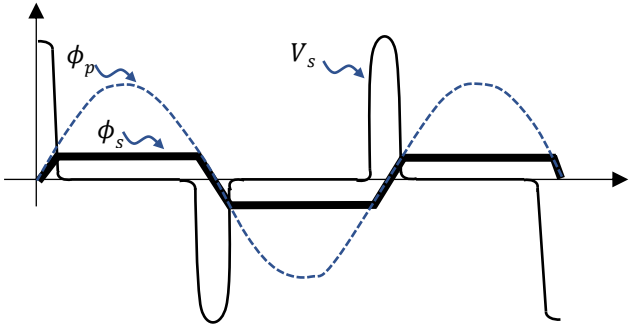

Fig. 20 Schematic of the flux in primary and secondary core legs and the secondary voltage of the peaking transformer.

$V_p = 100 \cos(2 \times \pi \times 60 \times t)$

According to Faraday's law, $\Phi_p = \frac{1}{N_p} \int V_p(t)\, dt$. Therefore, the flux in the primary winding can be calculated as follows:

$\Phi_p = \frac{100}{2 \times \pi \times 60 \times 1000} \sin(2 \times \pi \times 60 \times t) = 2.65\text{e-}04\ \sin(2 \times \pi \times 60 \times t)$

For the sake of simplicity, the hysteresis loop is ignored. According to the reluctance of the transformer core, which can be calculated using the provided dimensions in Fig. 19, the flux in the secondary core leg is as follows:

$\Phi_s = \frac{0.9954 \times 100}{2 \times \pi \times 60 \times 1000} \sin(2 \times \pi \times 60 \times t) = 2.64\text{e-}04 \sin(2 \times \pi \times 60 \times t)$

However, once the flux reaches $\phi_{sat,s} = B_{sat} \times A_s = 0.7 \times 4 \times 10^{-5} = 2.8 \times 10^{-5}$ at the secondary core leg of the transformer, which has a smaller cross section area, the secondary core leg path saturates and the remaining flux passes through the shunt leg. In this way, while the secondary core leg is saturated, the primary core leg is not saturated and the current and voltage at the primary winding remain sinusoidal.

Fig. 20 shows the schematic of the flux in the primary and secondary core legs according to above discussions. Also $V_s$ is determined using Faraday's law $V_s = N_s \frac{d\Phi_s}{dt}$. As shown in Fig.20, the above peaking transformer uses the nonlinearity of the transformer magnetization curve (i.e. saturation region of the transformer core) to generate voltage pulses $V_s$ in the output of the transformer.

## IV. CONCLUSION

This paper presented a unified approach using magnetic circuits to teach the equivalent electrical model of power transformers to undergraduate students. The discussed approach provides a direct connection between different elements of the electrical model and the real power transformers. Moreover, three examples presented that described how the behavior of power transformers in real-world applications can be explained using magnetic circuits.

## REFERENCES

[1] S. J. Chapman, "Electric Machinery Fundamentals" McGraw-Hill Education, 4th edition, 2001.
[2] S. Umans "Fitzgerald and Kingsley's Electric Machinery" McGraw-Hill Education, 7th Edition, 2013.
[3] T. Wildi, "Electrical Machines, Drives and Power Systems" Pearson, 6th Edition, 2014.
[4] A. Bergen, and V. Vittal "Power Systems Analysis" Pearson, 2nd Edition, 1999.
[5] R. W. Erickson, D. Maksimovic, "Fundamentals of Power Electronics" Springer, 3nd Edition, 2020.
[6] K. A. Macfadyen, "Vector Permeability" Journal of IEE, vol. 94, Pt. III, pp. 407–414, 1947.
[7] E. C. Cherry, "The Duality between Interlinked Electric and Magnetic Circuits and the Formation of Transformer Equivalent Circuits" Proceedings of the Physical Society. Section B, vol. 62, no. 2, pp. 101-111, 1949.
[8] P. R. Wilson, J. N. Ross, and A. D. Brown, "Modeling Frequency-Dependent Losses in Ferrite Cores" IEEE Transactions on Magnetics, vol. 40, no. 3, pp. 1537-1541, 2004.
[9] D. Buecherl, and H. G. Herzog "Iron Loss Modeling by Complex Inductances for Steady State Simulation of Electrical Machines" International Symposium on Power Electronics, Electric Drives, Automation and Motion, pp.878-883, 2010.
[10] A. P. S. Baghel, and S. V. Kulkarni "Modeling of Magnetic Characteristics Including Hysteresis Effects for Transformers" 3rd International Colloquium Transformer Research and Asset Management, Croatia, pp. 1-11, 2014.
[11] H. Tavakoli1, D. Bormann, D. Ribbenfjard, and G. Engdahl, "Comparison of a Simple and a Detailed Model of Magnetic Hysteresis with Measurements on Electrical Steel" The international journal for computation and mathematics in electrical and electronic engineering, vol. 28, no.3, pp. 1-5, 2009.
[12] S. Ayasun, and C. O. Nwankpa, "Induction Motor Tests Using MATLAB/Simulink and Their Integration Into Undergraduate Electric Machinery Courses" IEEE Transactions on Education, vol. 48, no. 1, pp. 37-46, 2005.
[13] Z. Sun, A. Schnabel, M. Burghoff, and L. Li "Calculation of an optimized design of magnetic shields with integrated demagnetization coils" American Institute of Physic ADVANCES, no.6, pp.1-9, 2016.
[14] E. Osmanbasic, and K. Obarcanin "Efficient Method to Accelerate Resistance Measurement of Transformer LV Winding" Transformers Magazine, vol. 5, no. 3, pp. 36-40, 2018.
[15] B. Hembroff, M. Ohlen, and P. Werelius, "A Guide to Transformer Winding Resistance Measurements" Megger Application Information, 2010.
[16] A. B. Thomas, "The Design of a Peaking Transformer" Journal of the British Institution of Radio Engineers, vol.13, no. 10, pp. 486-489, 1953.
[17] R. Lee, L. Wilson, and C. E. Carter, "Electronic Transformers and Circuits" Wiley-Interscience, 1988.

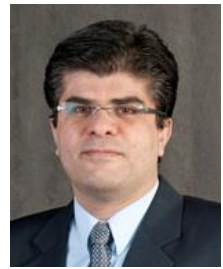

**Saeed Lotfifard** (S'08–M'11-SM'17) received his Ph.D. degree in electrical engineering from Texas A&M University, College Station, TX, in 2011. Currently, he is an associate professor at Washington State University, Pullman. His research interests include protection, control and operational security of inverter-based power grids. Dr. Lotfifard is an Editor for the IEEE Transaction on Smart Grid and IEEE Transactions on Sustainable Energy.